\documentclass[sigconf]{acmart}
\usepackage{longfbox}
\usepackage{graphicx}
\usepackage{amsmath}
\usepackage[american]{babel}
\usepackage{microtype}
\usepackage{subcaption}
\usepackage[ruled,linesnumbered ]{algorithm2e}
\usepackage[most]{tcolorbox}
\usepackage{balance}
\usepackage{enumitem} 
\usepackage{color}
\usepackage{url}
\usepackage{arydshln}
\usepackage{bm}
\usepackage{euscript}
\usepackage{mathrsfs}
\usepackage{tabularray}
\usepackage{xspace}
\usepackage{makecell}
\usepackage{bm}
\usepackage{multirow} 

\newcommand{\system}{\textsc{UICopilot}\xspace}%
\newcommand{\systemnospace}{\textsc{UICopilot}}%

\newcommand\mypara[1]{\vspace{1mm}\noindent \textbf{#1} \xspace}

\AtBeginDocument{%
  }

\copyrightyear{2025}
\acmYear{2025}
\setcopyright{acmlicensed}\acmConference[WWW '25]{Proceedings of the ACM
Web Conference 2025}{April 28-May 2, 2025}{Sydney, NSW, Australia}
\acmBooktitle{Proceedings of the ACM Web Conference 2025 (WWW '25), April
28-May 2, 2025, Sydney, NSW, Australia}
\acmDOI{10.1145/3696410.3714891}
\acmISBN{979-8-4007-1274-6/25/04}

\author{Yi Gui}
\authornote{These authors contribute equally to this research.}
\orcid{0009-0006-2841-7942}
\affiliation{%
  \institution{Huazhong University of Science and Technology}
  \city{Wuhan}
  \country{China}
}
\additionalaffiliation{%
\department{National Engineering Research Center for Big Data Technology and System, Services Computing Technology and System Lab, Cluster and Grid Computing Lab}
\department{School of Computer Science and Technology}
  \city{Wuhan}
  \state{Hubei}
  \country{China}}

\author{Zhen Li}
\authornotemark[1]
\authornotemark[2]
\orcid{0009-0007-0873-6126}
\affiliation{%
  \institution{Huazhong University of Science and Technology}
  \city{Wuhan}
  \country{China}
}

\author{Zhongyi Zhang}
\authornotemark[1]
\orcid{0009-0009-9951-3335}
\affiliation{%
  \institution{Huazhong University of Science and Technology}
  \city{Wuhan}
  \country{China}
}

\author{Yao Wan}
\authornote{Yao Wan is the corresponding author (\texttt{wanyao@hust.edu.cn}).}
\authornotemark[2]
\orcid{0000-0001-6937-4180}
\affiliation{%
  \institution{Huazhong University of Science and Technology}
  \city{Wuhan}
  \country{China}
}

\author{Dongping Chen}
\orcid{0009-0009-9848-2557}
\affiliation{%
  \institution{Huazhong University of Science and Technology}
  \city{Wuhan}
  \country{China}
}

\author{Hongyu Zhang}
\orcid{0000-0002-3063-9425}
\affiliation{%
  \institution{Chongqing University}
  \city{Chongqing}
  \country{China}
}

\author{Yi Su}
\orcid{0009-0007-9336-3562}
\affiliation{%
  \institution{Hubei University of Automotive Technology}
  \city{Shiyan}
  \country{China}
}

\author{Bohua Chen}
\orcid{0009-0004-8927-8902}
\affiliation{%
  \institution{Huazhong University of Science and Technology}
  \city{Wuhan}
  \country{China}
}

\author{Xing Zhou}
\orcid{0009-0006-4165-7756}
\affiliation{%
  \institution{Rabbitpre AI}
  \city{Shenzhen}
  \country{China}
}

\author{Wenbin Jiang}
\authornotemark[2]
\orcid{0000-0001-5628-8806}
\affiliation{%
  \institution{Huazhong University of Science and Technology}
  \city{Wuhan}
  \country{China}
}

\author{Xiangliang Zhang}
\orcid{0000-0002-3574-5665}
\affiliation{%
  \institution{University of Notre Dame}
  \city{Notre Dame}
  \country{United States}
}

\renewcommand{\shortauthors}{Yi Gui et al.}

\begin{document}

\title{\system: Automating UI Synthesis via Hierarchical Code Generation from Webpage Designs}

\begin{abstract}
Automating the synthesis of \textit{User Interfaces} (UIs) plays a crucial role in enhancing productivity and accelerating the development lifecycle, reducing both development time and manual effort.
Recently, the rapid development of \textit{Multimodal Large Language Models} (MLLMs) has made it possible to generate front-end \textit{Hypertext Markup Language} (HTML) code directly from webpage designs. 
However, real-world webpages encompass not only a diverse array of HTML tags but also complex stylesheets, resulting in significantly lengthy code. 
The lengthy code poses challenges for  
the performance and efficiency of MLLMs, especially in capturing the structural information of UI designs.
To address these challenges, this paper proposes \systemnospace, a novel approach to automating UI synthesis via hierarchical code generation from webpage designs.
The core idea of \system is to decompose the generation process into two stages: first, generating the coarse-grained HTML hierarchical structure, followed by the generation of fine-grained code.
To validate the effectiveness of \systemnospace, we conduct experiments on a real-world dataset, i.e., WebCode2M.
Experimental results demonstrate that \system significantly outperforms existing baselines in both automatic evaluation metrics and human evaluations.
Specifically, statistical analysis reveals that the majority of human annotators prefer the webpages generated by \system
over those produced by GPT-4V.\footnote{All the materials, including the source code and dataset, are available at: \url{https://github.com/CGCL-codes/naturalcc/tree/main/examples/uicopilot}.}
\end{abstract}



\maketitle

\section{Introduction}
Directly translating webpage designs into code significantly streamlines front-end development, reducing both development time and manual effort. This automation not only enhances developer productivity but also minimizes the risk of human error in code generation.
To this end, several approaches have been explored. pix2code~\cite{Tony2018_pix2code} was the first to generate \textit{Domain-Specific Language} (DSL) code from webpage designs by training a neural network, validating its effectiveness on a synthetic dataset. The generated DSL code could then be compiled into multiple front-end languages, including \textit{Hypertext Markup Language} (HTML). Similarly, Sketch2code~\cite{Alex2019_Sketch2code} leveraged deep learning and computer vision techniques to generate code from hand-drawn webpage sketches, further demonstrating the feasibility of automating design-to-code translation.

Recently, with the rapid advancement of \textit{Multimodal Large Language Models} (MLLMs), generating high-quality UI code from webpage designs (e.g., screenshots) has become increasingly feasible. Building on this progress, several exploratory studies have been conducted, focusing on dataset creation and benchmarking~\cite{Laurenccon2024UnlockingTC, gui2024vision2ui, Si2024Design2CodeHF, iwbench_guo_2024}. For example, WebSight~\cite{Laurenccon2024UnlockingTC} introduces a training dataset generated by an LLM, while Design2Code~\cite{Si2024Design2CodeHF} offers a curated test dataset of 485 samples along with an automatic metric to assess the similarity between generated webpages and their original designs. WebCode2M~\cite{gui2024vision2ui} provides a large-scale real-world dataset for both training and evaluation of webpage generation. Similarly, Web2Code~\cite{DBLP:journals/corr/abs-2406-20098} presents a large-scale synthesized dataset and an MLLM-based evaluation framework. When datasets are available, these works also fine-tune MLLMs to better fit the task of design-to-code translation.

\mypara{Challenges and Motivation.}
Despite the promising performance of the one-step generation approach, we are still far from fully automated UI code synthesis for real-world webpages. 
As noted in~\cite{Si2024Design2CodeHF}, the complexity of code generation increases significantly with the total number of HTML tags, the diversity of unique tags, and the depth of the \textit{Document Object Model} (DOM) tree.
When handling real-world webpage designs with intricate structures and a high number of unique HTML tags, existing MLLMs often suffer from notable declines in both performance and efficiency~\cite{Si2024Design2CodeHF}.
Specifically, the one-step generation approach faces two primary challenges.

\noindent\textbf{C1: The substantial length of the code that needs to be generated.}
From our investigation, existing code generation tasks mainly focus on generating short code snippets or functions with fewer than a few hundred tokens.
In contrast, real-world webpages consist not only of HTML but also complex \textit{Cascading Style Sheets} (CSS), significantly increasing the overall code length. For instance, webpages in the Common Crawl dataset often contain tens of thousands of tokens, and even after extensive cleansing, they still average over 5,000 tokens~\cite{gui2024vision2ui}. This far exceeds the context window of most existing LLMs, posing substantial challenges for both training and inference. As a result, webpage generation resembles project-level development rather than the straightforward generation of isolated code snippets.

\noindent\textbf{C2: The complexity of generating deeply nested structures.}
Furthermore, we observe that webpages typically consist of multiple layers of nested elements, making it particularly challenging to generate these intricate structures from high-resolution design diagrams. Prior studies~\cite{Kevin2020_MachineGUI, gui2024vision2ui} have demonstrated that even GPT-4V struggles to accurately capture structural information, as evidenced by its performance on the pix2code test dataset—one of the simplest benchmarks for code generation from webpage designs.

\begin{figure}[!t]
    \centering
    \includegraphics[width=0.98\linewidth]{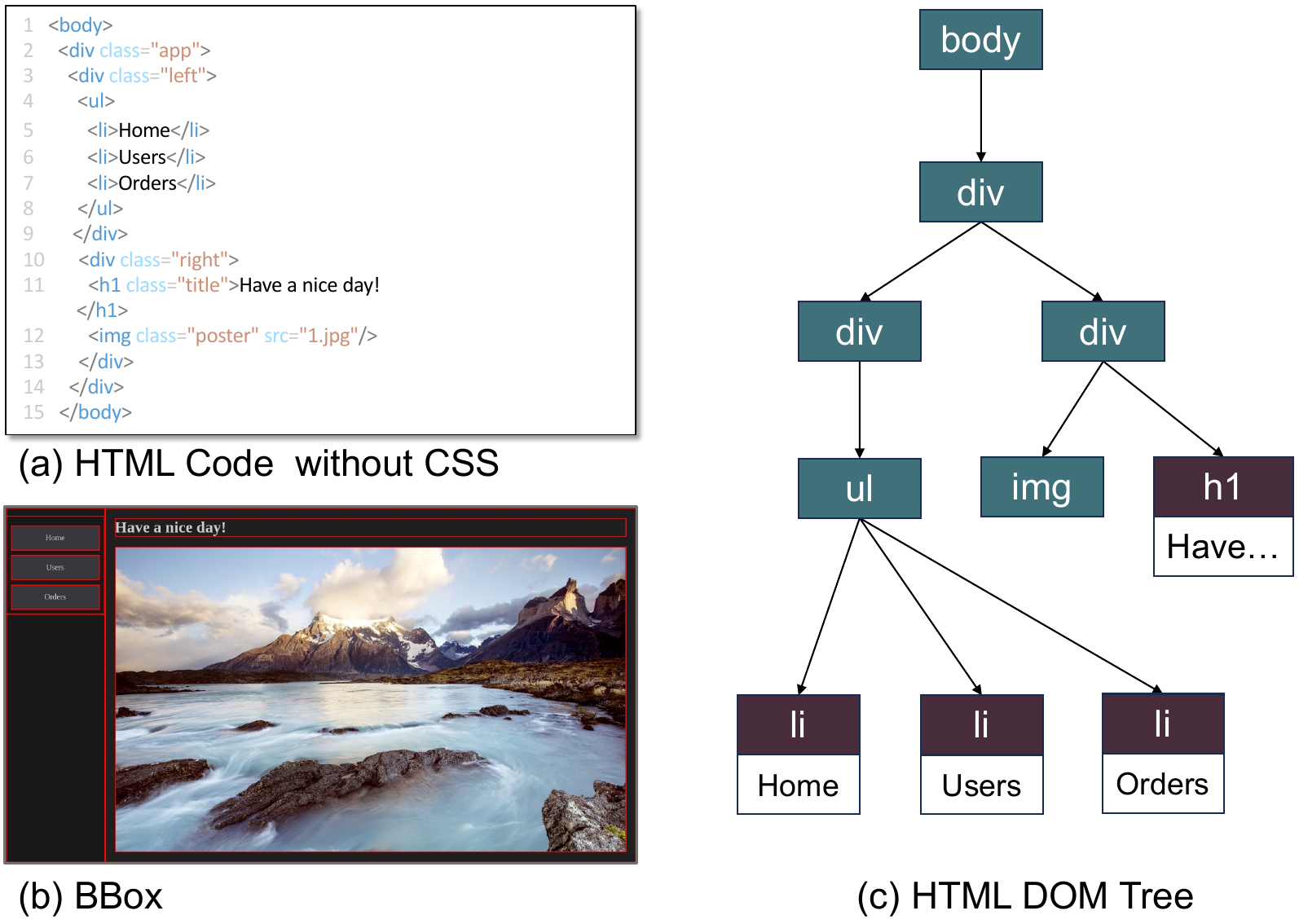}
    \caption{The structural information of a webpage.}
    \label{fig_structure}
\end{figure}

\begin{figure*}[!t]
    \centering
    \includegraphics[width=.98\textwidth]{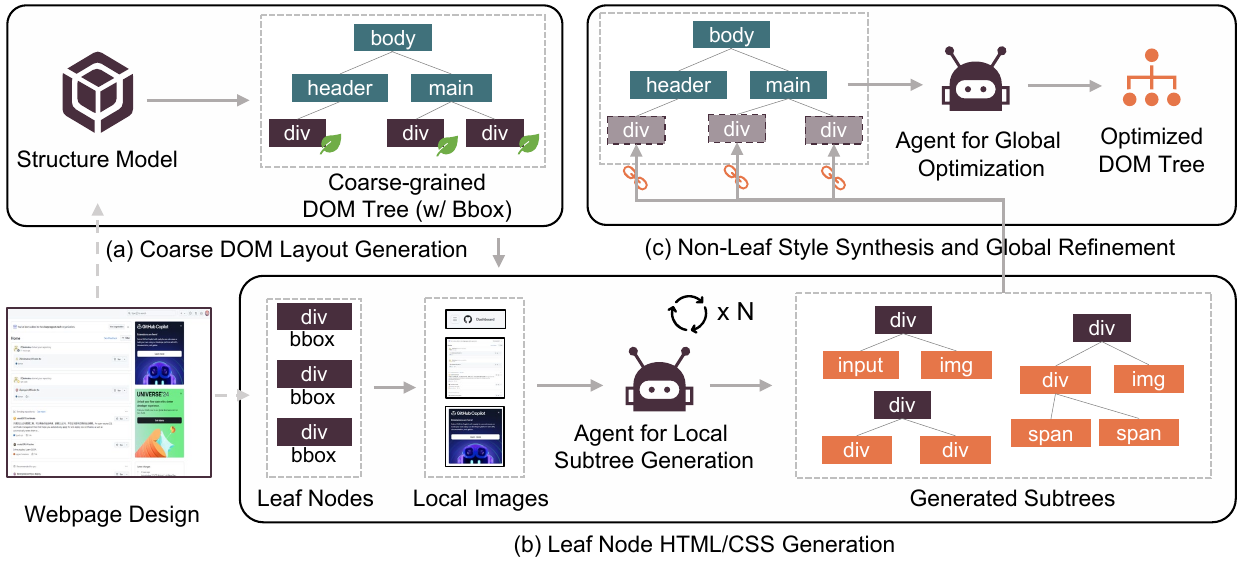}
    \caption{An overview of our proposed \systemnospace, which is composed of three components: (a) coarse DOM layout generation, (b) leaf node HTML/CSS generation, and (c) non-leaf style synthesis and global refinement.}
    \label{fig_overview}
\end{figure*}

\mypara{Our Work.} 
To tackle these challenges, this paper proposes \systemnospace, a novel approach for automating UI synthesis through hierarchical code generation from webpage designs (screenshots). To mitigate the complexity of generating lengthy code, we decouple the process into two stages: first, generating the coarse-grained hierarchical structure of HTML, and then producing the fine-grained code.
Specifically, we introduce and train a \textit{Vision Transformer} (ViT)-based structure model~\cite{alexey2020image} to predict a coarse DOM tree that captures only node types, hierarchy, and bounding boxes (BBoxes). To streamline training, we employ a BBox-based pruning strategy that simplifies the dataset and reduces prediction complexity by removing HTML tags with a BBox area below a fixed threshold, while retaining only node type and hierarchy information.
Using the predicted BBoxes of the coarse DOM tree’s leaf nodes, the original design image is segmented into subregion images. These subregions are then sequentially processed by a code generation agent, which generates the corresponding HTML/CSS code and embeds it back into the DOM tree. Finally, the DOM tree and design image are fed into the code generation agent again to refine the global code while generating styles and attributes for the non-leaf nodes.

To assess the effectiveness of \systemnospace, we conduct experiments on the real-world WebCode2M dataset, comparing it against several state-of-the-art baselines using three automatic metrics: CLIP~\cite{DBLP:conf/icml/RadfordKHRGASAM21}, SSIM~\cite{wang2004image}, and visual score~\cite{Si2024Design2CodeHF}.
The experimental results show that \system significantly outperforms other baselines. Notably, the visual score of GPT-4V improves by 23\%, 27\%, and 48\% when integrated into \system across three test datasets of increasing complexity, demonstrating that \system is particularly effective in handling complex webpage generation.
Additionally, we present the webpages generated by our method and GPT-4V in randomized order to human annotators for evaluation. The results reveal that in over 60\% of the cases, annotators prefer the webpages generated by our method, providing strong evidence for the effectiveness of \systemnospace.

The primary contributions of this paper are as follows:
\begin{itemize}[leftmargin=4mm, itemsep=0.05mm] 
\item We propose a novel approach \system 
that decouples the generation of hierarchical structure and fine-grained code. To the best of our knowledge, we are the first to break the limitations of one-step webpage code generation by addressing the challenges of lengthy code generation and complex nested structures in real-world webpage synthesis. 
\item We perform extensive experiments to evaluate the performance of  \system on the real-world WebCode2M dataset, and compare it with several state-of-the-art baselines. The results of automatic metrics and human evaluation show that \system consistently outperforms other baselines. 
\end{itemize}

\section{Preliminaries}
\subsection{Hierarchy Structure of Webpage Designs}
\label{sec_capture_structure}
Typically, webpage code consists of three main components: HTML, CSS, and JavaScript. In this work, we focus on generating static webpages and therefore exclude JavaScript.
The structural information in a webpage design diagram primarily comprises the hierarchical relationships between elements, along with their size and position. This information is crucial for ensuring high-quality webpage code generation and accurate final rendering.
First, webpage code follows a hierarchical, nested structure, represented by the HTML DOM tree (Figure~\ref{fig_structure}(c)), which serves as the backbone of the webpage. Second, the size and position of webpage elements are captured by \textbf{bounding boxes} (\textbf{BBox}) (Figure~\ref{fig_structure}(b)), which define the primary layout structure.
Our goal is to emphasize the preservation of a webpage’s hierarchical structure to generate code that closely aligns with the original design.

\subsection{Image to Code by MLLMs}
Recently, MLLMs have made significant progress in understanding and generating multimodal content, including text~\cite{DBLP:conf/aaai/FuSZY24}, images~\cite{DBLP:conf/iclr/PodellELBDMPR24}, and audio~\cite{DBLP:conf/icml/EvansCTHP24}.
In our scenario, we introduce two MLLMs: Pix2Struct-1.3B\cite{Kenton2023_Pix2Struct}, which serves as the structural model for hierarchical structure prediction, and GPT-4V\cite{DBLP:journals/corr/abs-2303-08774}, which acts as the code generation agent. We select Pix2Struct as our base MLLM due to its ViT-based image encoder and its aspect-ratio-preserving scaling strategy, making it more robust to extreme aspect ratios and adaptable to varying sequence lengths and resolutions.
Given an image, Pix2Struct first applies aspect-ratio-preserving scaling before splitting the image into fixed-size patches. These patches are then embedded and processed by the ViT encoder, which employs a self-attention mechanism to capture relationships between patches and understand the overall spatial arrangement of UI components. Finally, the model's Transformer-based text decoder takes the encoded representation and generates structured outputs, such as UI component expressions, functions, and locations.

\section{\system}
\subsection{Task Description}
Given a high-resolution webpage design, such as a screenshot of an existing webpage, \system aims to automatically generate the corresponding HTML and CSS code. The resulting webpage, once rendered, should closely resemble the input design in terms of layout, styling, and content. A crucial aspect of this process is accurately defining the structural elements of the page, including the type, dimensions, and positioning of components, as these factors directly influence the visual integrity of the generated webpage. Consequently, our work prioritizes the precise synthesis of structured and well-formed webpage layouts to ensure fidelity to the original design.

\subsection{Overview}
As shown in Figure~\ref{fig_overview}, we decouple the webpage generation process into two stages: coarse DOM layout generation (Figure~\ref{fig_overview}(a)) and fine-grained code synthesis (Figure~\ref{fig_overview}(b\&c)).
Our approach is based on the assumption that ``\textit{when the elements and structure in a design image are sufficiently simple, MLLMs can effectively generate webpage code that captures both structure and detail}''.
By separating the generation of the DOM tree and positional information from the detailed local code, our approach alleviates the burden on LLMs when handling lengthy code generation. Additionally, prioritizing the DOM tree and positional details enhances the model's ability to accurately preserve the original webpage's structural elements, effectively addressing the challenges of generating lengthy code and capturing structural information, as mentioned before.

\begin{figure}[!t]
    \centering
    \includegraphics[width=0.94\linewidth]{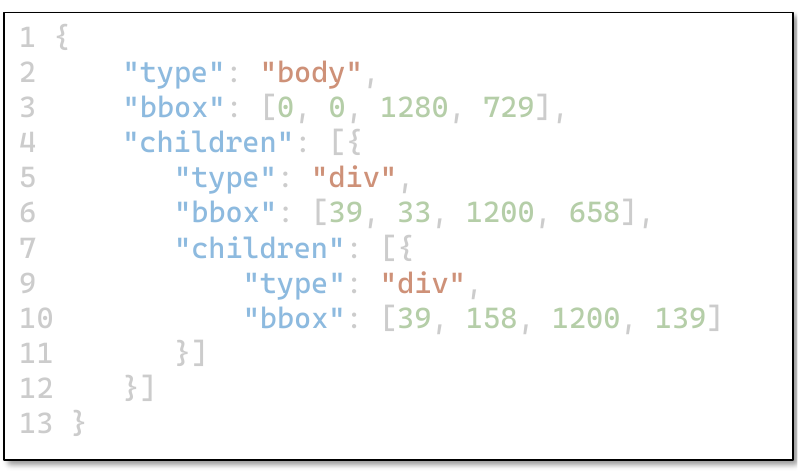}
    \caption{An example of DOM tree in JSON format.}
    \label{fig_json}
\end{figure}

\subsection{Coarse DOM Layout Generation}
\label{sec_dom_generation}
To generate webpage structure information more accurately and efficiently, we introduce and train a structure model specifically designed to predict a coarse-grained version of the HTML DOM tree, capturing only the node types, hierarchy, and BBoxes of the nodes.
Additionally, we carefully design the data pruning and training procedures to optimize the performance.

\mypara{Structure Model.}
In our scenario, high-resolution webpage designs appear in various resolutions, and directly cropping or resizing them often results in the loss of critical details, negatively impacting UI code generation. To address this, we draw inspiration from ViT’s approach of segmenting images into multiple patches to retain as much information as possible. Based on this principle, we adopt Pix2Struct, a ViT-based model, as our core structure model. Pix2Struct is particularly well-suited for this task due to its robustness to extreme aspect ratios and its ability to dynamically adapt to varying sequence lengths and resolutions.

We formulate the structure prediction task as a sequence generation problem, following the next-token prediction paradigm. Given an input image \( I \), which is a webpage design, the model produces a structured representation of the webpage's DOM tree in JSON format (exemplified in Figure~\ref{fig_json}). This JSON text encodes the DOM hierarchy as nested HTML elements, along with corresponding BBox attributes.  
The process begins by segmenting the input image \( I \) into fixed-size patches, represented as \( \{P_1, P_2, \dots, P_N\} \). These patches are then processed by the ViT encoder, \( \text{Enc}_{\theta} \), which transforms them into hidden state vectors as follows:
\begin{equation}
\{h_1, h_2, \dots, h_N\} = \text{Enc}_{\theta}(\{P_1, P_2, \dots, P_N\})\,.
\end{equation}
These hidden state vectors are subsequently fed into the decoder, \( \text{Dec}_{\phi} \), a conventional Transformer-based text decoder, to predict the JSON representation of the DOM tree and their corresponding BBoxes. Leveraging the encoder's hidden states, the decoder generates the JSON text one token at a time. At each time step \( t \), the decoder generates a probability distribution for the next token \( y_t \), conditioned on all previous tokens \( y_1, y_2, \dots, y_{t-1} \). This prediction is computed using a softmax function applied to the decoder’s hidden state \( h_t \), which integrates information from both the encoder's hidden states \( \{h_1, h_2, \dots, h_N\} \) and the previously generated tokens, projected by a learned weight matrix \( W \) as follows:
\begin{equation}
P(y_t \mid y_1, y_2, \dots, y_{t-1}) = \text{softmax}(W \cdot h_t)\,.
\end{equation}

\mypara{BBox-based Data Pruning.}
\begin{figure}[t!]
    \centering
    \includegraphics[width=0.98\linewidth]{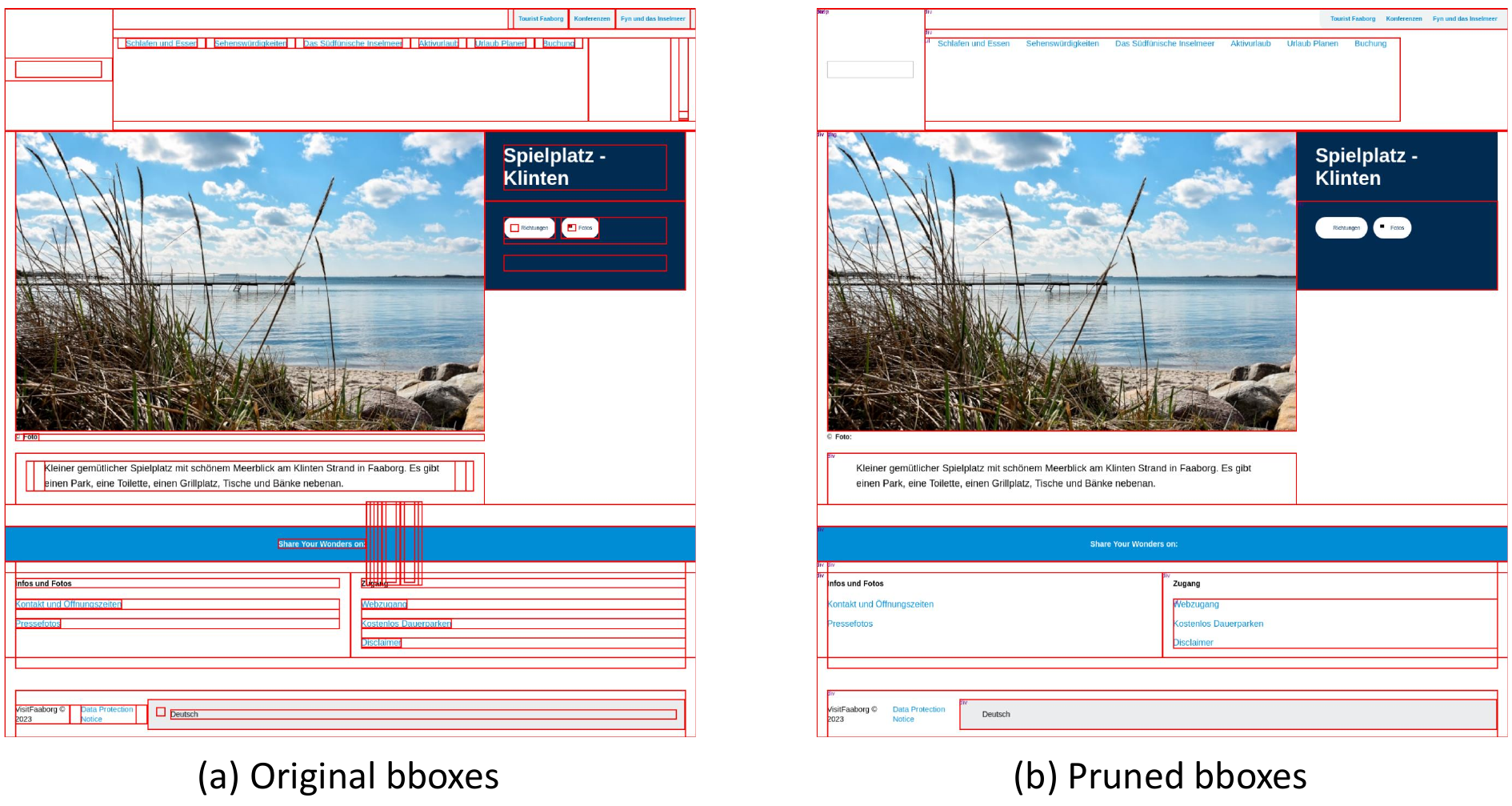}
    \caption{BBox-based data pruning.}
    \label{fig_bbox_prune}
\end{figure}
Throughout our investigation, we find that the BBoxes of HTML elements in training datasets often contain a significant amount of noise, such as empty or very small elements, as well as elements representing visually hidden parts of the webpages or parts that are not displayed correctly (as illustrated in Figure~\ref{fig_bbox_prune}(a)). 
The noise not only severely reduces the model's learning efficiency but also contributes little to the overall structural information. 
Therefore, we apply several heuristic rules to prune the original BBoxes along with their corresponding webpage elements:
\textbf{(1)} We first remove BBoxes smaller than 3\% of the total area, along with all their child nodes. We reasonably assume that for simple content within small areas, the MLLM can efficiently generate the corresponding high-quality structure and style code.
\textbf{(2)} We eliminate BBoxes that contain only solid-colored pixels, along with all their child nodes. These BBoxes are usually empty and contain no actual web elements.
\textbf{(3)} We discard webpage samples with fewer than 10 total BBoxes on the entire page, since an insufficient number of BBoxes typically indicates potential errors during processing.

Figure~\ref{fig_bbox_prune}(b) presents an example of the pruned BBoxes, where most of the noise has been eliminated, revealing a clearer and simplified hierarchical structure. During model training, we find that this pruning method significantly accelerates model convergence while preserving as much of the original hierarchical structure of the webpage as possible. 
During the inference process of the structure model, this streamlined BBox information greatly improves both the generation quality and efficiency.
We also apply pruning rules to the results generated by the structure model, primarily based on minimum area (\texttt{min\_area}) of BBoxes and maximum depth (\texttt{max\_depth}) of DOM tree.
This ensures that the predicted DOM tree and BBoxes are concise and appropriate, setting a better stage for fine-grained code generation. 

\subsection{Fine-Grained Code Synthesis}
While the generated coarse DOM tree contains the node types and hierarchy, it omits two crucial aspects:
\textbf{(1)} The structural and style code for the local regions of the leaf nodes. Since the structure model is designed to predict a coarse-grained DOM tree, and this prediction is further pruned based on \textit{min\_area} and \textit{max\_depth}, the leaf nodes in the DOM tree actually serve as parent nodes for subregions that still contain sub-trees requiring further prediction.
\textbf{(2)} The attributes and styles of the non-leaf nodes within the coarse DOM tree.
To address these two missing elements, we have devised the following two steps to complete the generation process.

\begin{figure}[t!]
    \centering
    \includegraphics[width=\linewidth]{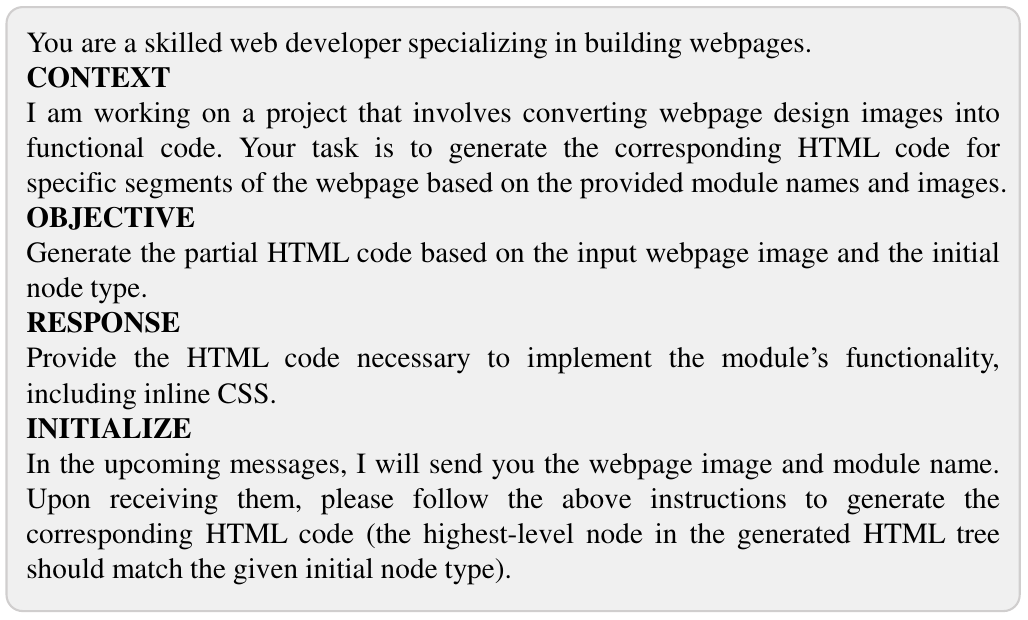}    
    \caption{The prompt for leaf node HTML/CSS generation.}
    \label{fig_prompts_1}
\end{figure}

\begin{figure}[t!]
    \centering
    \includegraphics[width=\linewidth]{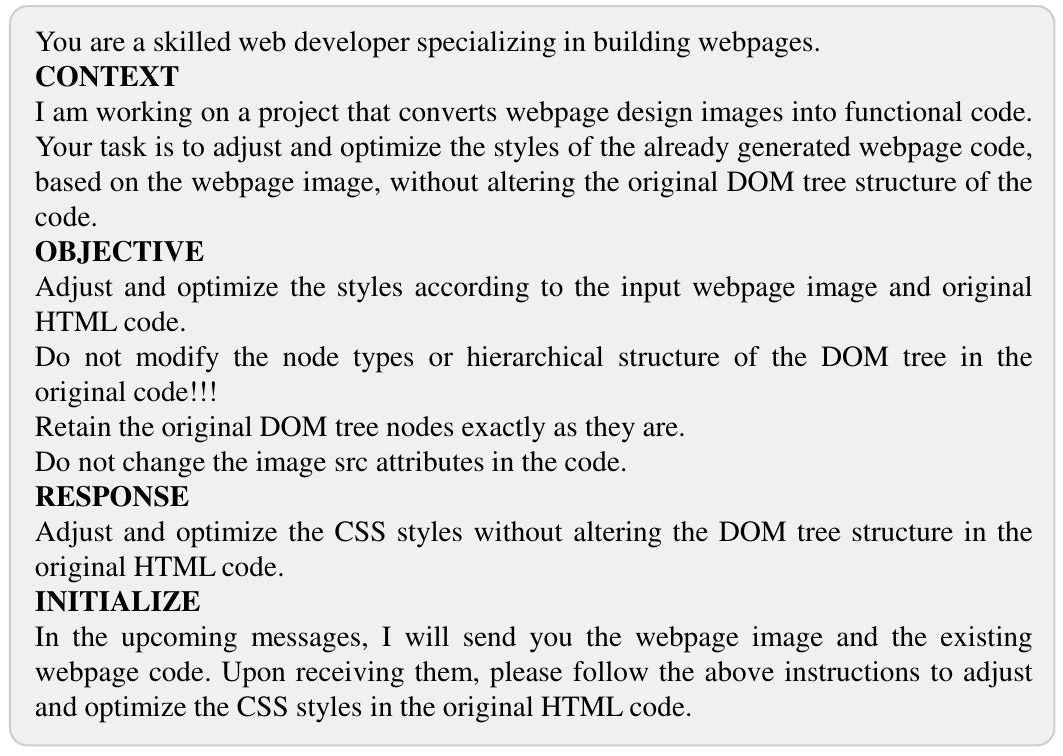}    
    \caption{The prompt for non-leaf style synthesis and global refinement.}
    \label{fig_prompts_2}
\end{figure}

\mypara{HTML/CSS Generation of Leaf Node.}
In practice, we employ GPT-4V as our local code generation agent, leveraging its flexibility and robust code generation capabilities.
Since the BBoxes contain the size and position information of the nodes, we use the BBoxes of the leaf nodes in the predicted coarse DOM tree to crop images of the corresponding subregions from the original image.
These segmented images are then fed individually into the code generation agent to predict the corresponding HTML/CSS code for each subregion. 
After obtaining the HTML/CSS code for the leaf nodes, we embed them back into the corresponding leaf nodes of the coarse DOM tree.
By doing so, we isolate the visual content associated with each leaf node, allowing the agent to generate accurate HTML/CSS code for these specific regions one by one, significantly alleviating the burden of generating lengthy code and improving the generation quality.
We have carefully designed a prompt that instructs the agent to generate the corresponding HTML/CSS code based on the subregion image and the parent node type, as shown in Figure~\ref{fig_prompts_1}.

\mypara{Non-Leaf Style Synthesis and Global Refinement.}
At this stage, we provide the code from the previous step along with the full design image to the agent, instructing it to add styles and other attributes to the non-leaf nodes of the coarse DOM tree. The agent analyzes the global layout and visual elements to infer styling details like fonts, colors, margins, paddings, and other CSS properties, ensuring that the final code reflects both the structural hierarchy and the visual aesthetics of the original design. Additionally, the agent refines the global code while preserving the input DOM tree. The prompt used in this process is shown in Figure~\ref{fig_prompts_2}.

\section{Experiments and Analysis}

\subsection{\textbf{Datasets}}
\mypara{Training Dataset.}
We utilize two datasets, WebSight v0.1 and WebCode2M, for training the structure model, each demonstrating distinct characteristics across various indicators.
As shown in Table~\ref{tb_dataset_stat}, compared to the WebSight dataset, WebCode2M's data is more complex, possesses a richer variety of styles and tag types, and is significantly longer, making it closer to real-world HTML code.
Furthermore, WebCode2M provides the page's BBox information directly, which is essential for the training of our structure model.
For the WebSight dataset, we also extract the BBox information prior to training.
\begin{table}[!t]
\centering
\caption{A statistical comparison between both WebSight and WebCode2M. The statistical data of the two is referred to~\cite{Si2024Design2CodeHF}.}
\begin{tabular}{l|cc}
\Xhline{1px}
                    & WebSight                        & WebCode2M   \\ 
\Xhline{0.7px}
Purpose             & Training                           & Training\&Tesing    \\
Source              & Synthetic   & Real-World (Common Crawl)      \\
Size                & 0.8M                                  & 3.1M               \\
Avg. Len (tokens) & 647±216                         & 4661±2006         \\
Avg. Tags      & 19±8                                & 188±80            \\
Avg. DOM Depth      & 5±1                                    & 15±5              \\
Avg. Unique Tags     & 10±3                            & 24±6              \\
\Xhline{1px}
\end{tabular}
\label{tb_dataset_stat}
\end{table}
\begin{table*}[!t]
\centering
\caption{The performance breakdown on the visual metrics.}
\small
\setlength{\tabcolsep}{4pt} 
\begin{tabular}{l|ccc|ccc|ccc} 
\Xhline{1px}
\multirow{2}{*}{Model} & \multicolumn{3}{c|}{WebCode2M-Short}                                  & \multicolumn{3}{c|}{WebCode2M-Mid}                                    & \multicolumn{3}{c}{WebCode2M-Long}                                                \\
                       & Visual Score        & CLIP             & SSIM           & Visual Score         & CLIP             & SSIM           & Visual Score         & CLIP             & SSIM                       \\ 
\Xhline{0.7px}
WebSight VLM-7B           & 0.57 (±0.24)          & 0.69 (±0.12)          & 0.62 (±0.17)          & 0.52 (±0.23)          & 0.67 (±0.11)          & 0.59 (±0.16)          & 0.48 (±0.27)          & 0.64 (±0.11)          & 0.61 (±0.15)  \\
Design2Code            & 0.75 (±0.14)          & 0.68 (±0.10)          & 0.58 (±0.15)          & 0.69 (±0.23)          & 0.70 (±0.10)          & 0.56 (±0.14)          & 0.61 (±0.28)          & 0.68 (±0.10)          & 0.61 (±0.11)                      \\ 

CogAgent-Chat          & 0.46 (±0.31)          & 0.68 (±0.11)          & 0.59 (±0.15)          & 0.40 (±0.31)          & 0.66 (±0.10)          & 0.58 (±0.14)          & 0.39 (±0.30)          & 0.65 (±0.10)          & 0.60 (±0.13)                      \\ 

LLaVA-v1.5-7B          & 0.43 (±0.27)          & 0.63 (±0.11)          & \textbf{0.65 (±0.17)}          & 0.21 (±0.28)          & 0.63 (±0.10)          & \textbf{0.65 (±0.14)} & 0.19 (±0.27)          & 0.61 (±0.10)          & \textbf{0.66 (±0.12)}             \\

GPT-4V                 & 0.68 (±0.32)          & 0.74 (±0.10)          & 0.61 (±0.14)          & 0.65 (±0.33)          & 0.71 (±0.10)          & 0.55 (±0.12)          & 0.62 (±0.35)          & 0.67 (±0.10)          & 0.57 (±0.11)                      \\ 

\system     & \textbf{0.84 (±0.18)}          & \textbf{0.77 (±0.11)}          & 0.60 (±0.13)          & \textbf{0.83 (±0.17)}          & \textbf{0.77 (±0.10)}          & 0.57 (±0.12)          & \textbf{0.78 (±0.24)}          & \textbf{0.74 (±0.10)}          & 0.60 (±0.11) \\
\Xhline{1px}
\end{tabular}
\label{tb_performance}
\end{table*}

\mypara{Test Dataset.}
We evaluate our framework on the WebCode2M test datasets.
WebCode2M test datasets are composed of three subsets: WebCode2M-Short, WebCode2M-Mid, and WebCode2M-Long. 
These subsets are created by splitting the data according to the length range of the ground-truth HTML code.
The length ranges of the ground-truth HTML code for these three subsets are [551, 2045], [2052,4085], and [4098,10990], respectively. Each subset contains 256 samples.

\subsection{Evaluation Metrics}

\mypara{CLIP Similarity~\cite{DBLP:conf/icml/RadfordKHRGASAM21}.}
CLIP is a multi-modal model trained using a contrastive objective on a large dataset of millions of internet text-image pairs. It learns to align images with their textual descriptions in a shared representation space. The latent vectors generated by the CLIP model capture the semantic information of the inputs. As a result, the cosine similarity between these vectors, which calculates the cosine of the angle between them, effectively quantifies the degree of similarity between the images.

\mypara{SSIM (\textit{Structural Similarity Index})~\cite{wang2004image}.}
It considers changes in structural, luminance, and contrast information in the images.

\mypara{Visual Score~\cite{Si2024Design2CodeHF}.}
It is used to assess the degree of alignment between low-level elements based on their appearance. The scores primarily evaluate the match ratio between reference and candidate blocks, as well as their similarity at the block level, considering factors such as color, text, and position.\footnote{We use the visual score from the first version of their paper, which may differ from the latest version, especially with regard to the sub-indicator \textit{text color}.}

\subsection{Baselines}
The baselines in our work fall into two categories: MLLMs specialized for webpage generation and general-purpose MLLMs. The specialized MLLMs include:
\begin{itemize}[leftmargin=4mm, itemsep=0.05mm] 
    
\item{\textbf{WebSight VLM-8B~\cite{Laurenccon2024UnlockingTC}.}}
Hugging Face's WebSight utilizes its training dataset and the DoRA~\cite{DBLP:journals/corr/abs-2402-09353} mechanism to fine-tune a VLM that has been pre-trained on image/text pairs. 

\item{\textbf{Design2Code-18B~\cite{Si2024Design2CodeHF}.}}
Stanford's Design2Code is also fine-tuned on the WebSight dataset. However, it adopts CogAgent~\cite{DBLP:journals/corr/abs-2312-08914} as its base model and utilizes LoRA~\cite{DBLP:conf/iclr/HuSWALWWC22} as the fine-tuning method to accelerate the training process. 
\end{itemize}

\noindent{The general-purpose (prompt-based) MLLMs include:}
\begin{itemize}[leftmargin=4mm, itemsep=0.05mm]     

\item{\textbf{CogAgent-Chat-18B~\cite{DBLP:journals/corr/abs-2312-08914}.}}
CogAgent-Chat-18B is a general MLLM that supports both low- and high-resolution images. Notably, it performs well on webpage navigation, requiring only screenshots. We input the screenshot and a simple prompt ``\textit{write an HTML code}'' to generate the webpage, similar to Design2Code.

\item{\textbf{GPT-4V~\cite{DBLP:journals/corr/abs-2303-08774}.}}
GPT-4V has shown remarkable capabilities in image comprehension. It also possesses the unique ability to generate code from images. We refer to the prompt of the well-known open-source project \textit{screenshot-to-code}~\footnote{https://github.com/abi/screenshot-to-code} on GitHub, with slight modifications. 

\item{\textbf{LLaVA-v1.5-7B~\cite{liu2023llava}.}}
LLaVA-v1.5-7B is an MLLM that connects a vision encoder and an LLM for general-purpose visual and language understanding. We use the same prompt as GPT-4V to generate webpages from images.
\end{itemize}

\subsection{The Effectiveness of \system}
\mypara{Overall Performance.} Table~\ref{tb_performance} presents the performance breakdown of \system and the baselines on the WebCode2M test datasets. From the table, we observe that our method's visual score and CLIP similarity significantly outperform all the baselines across the three test datasets. This leading performance in the CLIP metric indicates that the webpages generated by \system are more visually similar to the original ones in terms of overall appearance and features.
In terms of SSIM, LLaVA holds a slight advantage over the other models, suggesting that the webpages it generates may be more closely aligned with the originals concerning visual luminance, contrast, and structure. However, the differences among all models in this metric are not significant. Our \system also performs well on this metric compared to the other models.

Notably, while \system utilizes GPT-4V for generating fine-grained code, it still outperforms GPT-4V in terms of visual score by 23\%, 27\%, and 48\% across the three test datasets, respectively.
Additionally, \system exhibits significantly lower variance in visual scores, indicating higher stability. Moreover, it is worth noting that as the length of the sample webpages in the test datasets increases, the advantage of \system over GPT-4V grows substantially. This further demonstrates that \systemnospace's architecture—which focuses on decoupling the structure parsing from the fine-grained code generation processes—reduces complexity and results in more accurate HTML/CSS code generation from design images.

\begin{figure}[!t]
    \centering
    \includegraphics[width=0.98\linewidth]{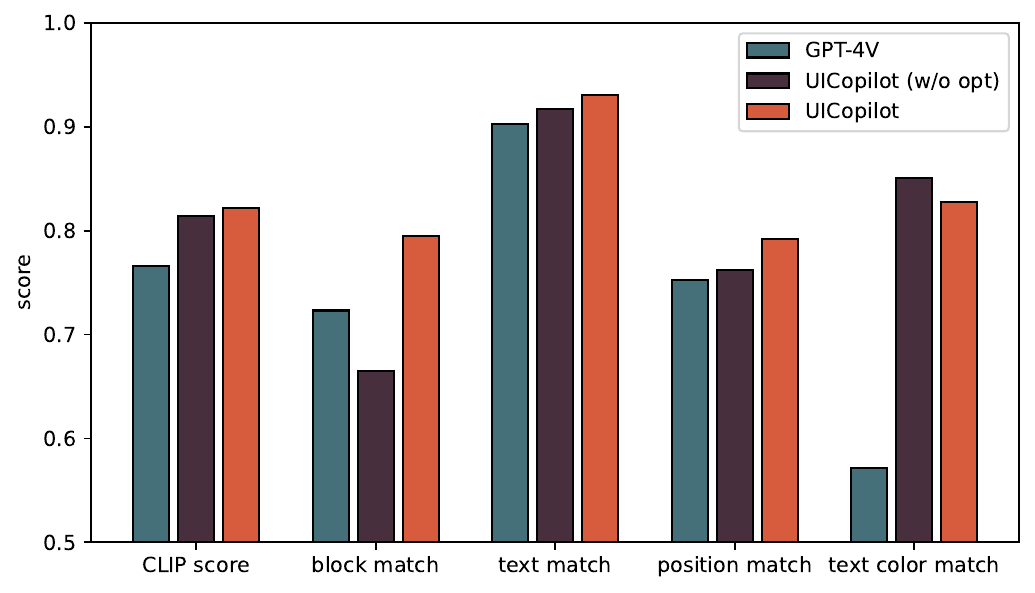}
    \caption{Detailed sub-indicators of visual score.}
    \label{fig_visual_scores}
\end{figure}

\mypara{Visual Score Breakdown.}
The visual score is a composite metric consisting of five sub-indicators: block-level color, text, position, text color, and CLIP similarity information. In Figure~\ref{fig_visual_scores}, we provide a detailed analysis of these sub-indicators.
To further explore the contribution of the refinement process in our framework, we introduce \system without the global refinement (denoted as \system w/o opt) for comparison. 
As shown in Figure~\ref{fig_visual_scores}, \system without the refinement already outperforms GPT-4V in terms of CLIP similarity, text, position, and text color.
This demonstrates that the input coarse-grained HTML DOM tree and BBoxes effectively enhance GPT-4V’s generation capability, particularly evident in the significant improvement in the text color indicator.
However, we observe that \system without refinement performs worse than GPT-4V in the block-level color indicator. Our investigation suggests that strictly adhering to the coarse-grained HTML DOM tree and BBoxes without refinement can lead to missing styles or errors in the upper layers of nodes, negatively impacting the block-level color. After incorporating the refinement process, as seen in Figure~\ref{fig_visual_scores}, there is a significant improvement in the block-level color indicator, highlighting the effectiveness of the refinement process.

\subsection{The Influence of \texttt{min\_area} and \texttt{max\_depth}}
\begin{figure}[!t]
    \centering
    \includegraphics[width=0.98\linewidth]{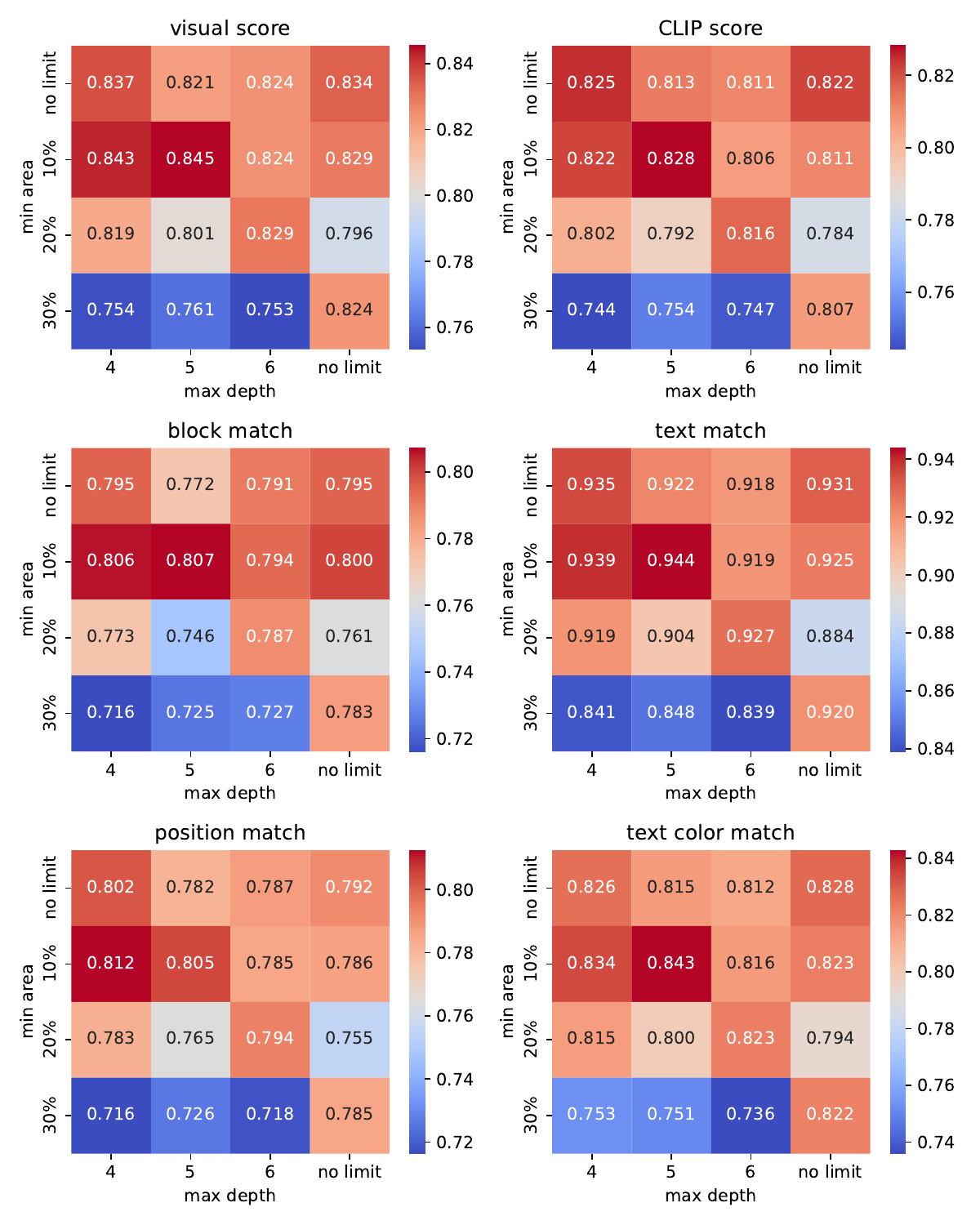}
    \caption{Grid search results of parameter optimization.}
    \label{fig_heatmaps}
\end{figure}

As described in Section~\ref{sec_dom_generation}, we prune the generated results by the structure model based on two parameters: the minimum area of BBox (\texttt{min\_area}) and the maximum depth of the DOM tree (\texttt{max\_depth}). To search for the optimal values of these parameters, we conduct a grid search to explore different combinations. However, since performing inference on the entire test dataset is time-consuming, we optimize the grid search process by constraining the search space based on empirical experience: [10\%, 20\%, 30\%, Unlimited] for \texttt{min\_area} and [4, 5, 6, Unlimited] for \texttt{max\_depth}.
We focus on the metric with the highest relative improvement in our method—the visual score—for further experiments. By analyzing its sub-metrics under different combinations of \texttt{min\_area} and \texttt{max\_depth}, we obtain the results shown in Figure~\ref{fig_heatmaps}.
Our analysis reveals a consistent linear trend across nearly all sub-metrics: as both \texttt{max\_depth} and \texttt{min\_area} decrease, the visual score improves, indicating enhanced visual quality in the generated results. Specifically, we find that setting \texttt{max\_depth} to 4 and \texttt{min\_area} to 10\% achieves a well-balanced local optimum for visual matching.

\subsection{Human Evaluation}
\begin{figure}[!t]
    \centering
    \includegraphics[width=0.98\linewidth]{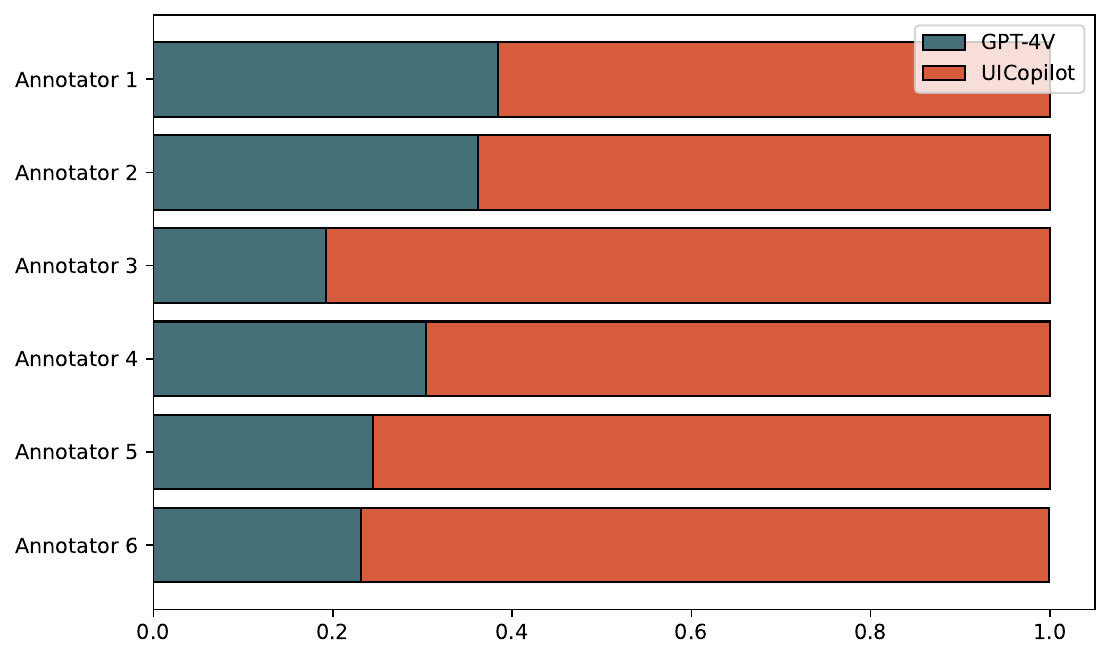}
    \caption{The preference of human annotators.}
    \label{human_preference}
\end{figure}

\begin{figure*}[!t]
    \centering
    \includegraphics[width=\linewidth]{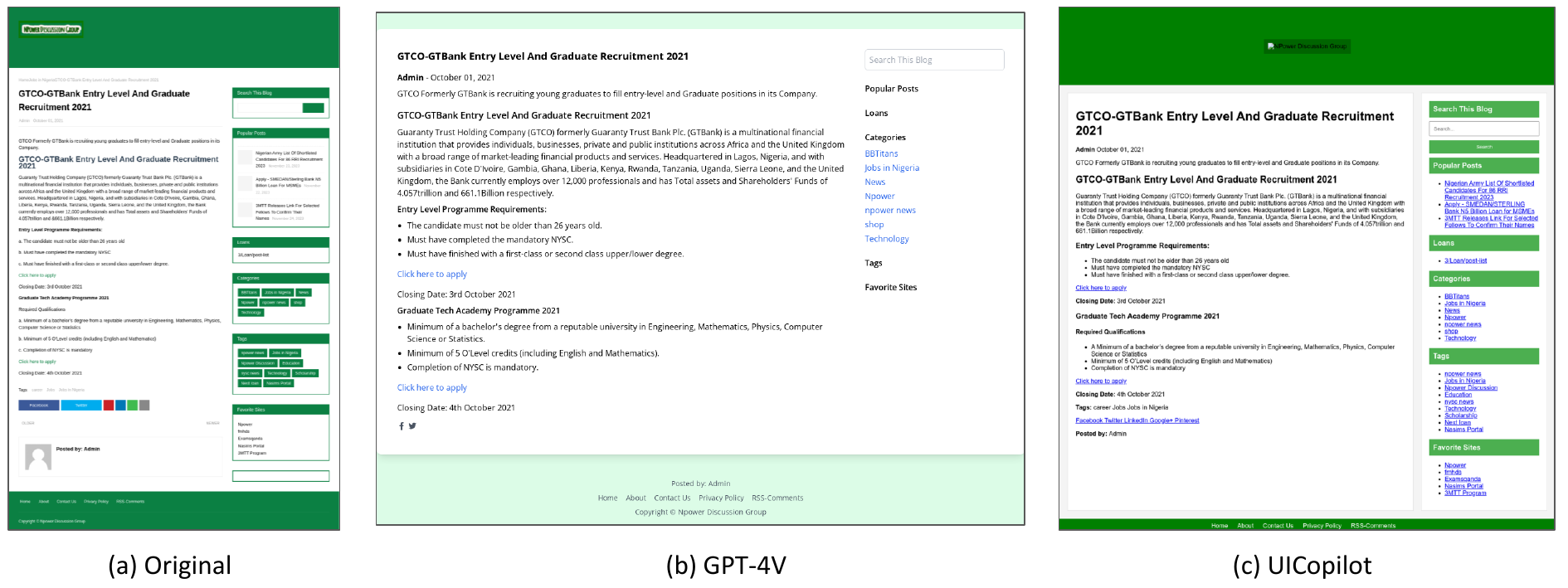}
    \caption{A representative example of the generated webpages by GPT-4V and \systemnospace.}
    \label{fig_case}    
\end{figure*}
From Table~\ref{tb_performance}, we observe minimal differences between \system and GPT-4V in terms of CLIP similarity and SSIM, both of which primarily measure overall image similarity. To address the question, ``\textit{which model generates webpages that are closer to the original design?}'', we conduct a human evaluation study. Specifically, we present six annotators with pairs of webpages generated by \system and GPT-4V, shuffling the samples in each pair to eliminate bias. Annotators are asked to select the webpage they find most aligned with the original design. As shown in Figure~\ref{human_preference}, over 60\% of selections favored \system, providing strong evidence of its effectiveness in generating webpages that more closely resemble the original designs.

\subsection{Case Study}
To highlight the advantages of \system in webpage generation, we present a representative example in Figure~\ref{fig_case}. The images, from left to right, show the original webpage, the one generated by GPT-4V, and the one generated by \system. Both GPT-4V and \system capture the footer, body, and header, with text content close to the original. However, GPT-4V simplifies the small blocks on the right side, while \system accurately replicates these structural details. This demonstrates that decomposing webpage synthesis into coarse-grained DOM tree generation and fine-grained localized code enhances structural accuracy. Despite this, both models still miss finer details such as text, colors, and section sizes, suggesting that post-generation edits and localized fixes could improve webpage quality, making them valuable areas for future exploration.

\section{Related Work}
\mypara{Image Representation Learning.}
Learning effective image representations is fundamental to a wide range of computer vision tasks, including image-to-text applications such as image captioning~\cite{DBLP:conf/cvpr/VinyalsTBE15}, image-to-code translation~\cite{Si2024Design2CodeHF}, image classification~\cite{DBLP:journals/ijcv/RussakovskyDSKS15}, object detection~\cite{DBLP:conf/cvpr/GirshickDDM14}, and semantic segmentation~\cite{DBLP:journals/pami/ShelhamerLD17}. Extensive research has explored various strategies for extracting meaningful latent representations from images. For instance, \citet{chen2020simple} leveraged contrastive learning to train image encoders on large-scale datasets, improving feature extraction through self-supervision. ViT~\cite{alexey2020image} introduced an innovative paradigm by segmenting images into fixed-size patches and processing them using a Transformer-based architecture, enabling adaptability to varying resolutions. More recently, diffusion models~\cite{DBLP:conf/nips/HoJA20,DBLP:conf/cvpr/RombachBLEO22} have emerged as powerful tools for image representation and visual data generation.

\mypara{Code Generation.}
Neural language models have seen remarkable progress in code intelligence~\cite{wan2024deep}, enabling advancements in key tasks such as code summarization~\cite{wan2018improving,wang2020reinforcement}, code search~\cite{wan2019multi}, and code generation~\cite{bi2024iterative,DBLP:conf/kbse/Sun000J0L24}. With the rapid evolution of LLMs in text generation, numerous code-specific LLMs have emerged, including CodeT5+~\cite{wang2023codet5+}, InCoder~\cite{fried2022incoder}, StarCoder~\cite{Li2023StarCoderMT}, Code Llama~\cite{roziere2023code}, WizardCoder~\cite{luo2023wizardcoder}, Qwen-Coder~\cite{hui2024qwen2}, and DeepSeek-Coder~\cite{guo2024deepseek}.
These models have revolutionized software development by powering production tools such as Copilot~\cite{GitHub2022copilot}, CodeWhisperer~\cite{Aws2022codewhisperer}, and Replit~\cite{Replit2022replit_ai}, streamlining coding workflows and enhancing developer productivity. While much research has focused on general-purpose LLMs for code generation, this paper explores a distinct challenge: generating code directly from webpage designs. This specialized task requires a nuanced understanding of UI structures, layout constraints, and code synthesis, setting it apart from traditional code generation paradigms.

\mypara{Image to Code.}
Several prior studies have focused on generating code directly from images. For instance, Sketch2Code~\cite{Alex2019_Sketch2code} generates website code from wireframe sketches using two complementary approaches: a computer vision-based method that identifies visual elements and structures, and a deep learning-based method that translates sketches into structured, normalized webpages suitable for code generation. Pix2Struct~\cite{Kenton2023_Pix2Struct}, pre-trained on predicting simplified HTML code from masked website screenshots, has demonstrated substantial improvements in visual-language understanding tasks, achieving state-of-the-art performance across nine tasks spanning four domains.
\citet{Wu2021ScreenPT} tackled the problem of screen parsing by predicting UI hierarchy graphs from screenshots. They utilized Faster-RCNN~\cite{Ren2015FasterRT} for image encoding and an LSTM-based attention mechanism to construct the graph nodes and edges.
Additionally, some studies focus on curating specialized training datasets~\cite{Laurenccon2024UnlockingTC, gui2024vision2ui} to improve MLLMs' capabilities in UI generation, while others aim to establish benchmarks and evaluation metrics~\cite{Si2024Design2CodeHF, xiao2024interaction2code, chen2024gui} for assessing these models.

\section{Conclusion}
In this paper, we propose \system, a novel approach to generating code from webpage designs by decomposing the process into two stages: coarse DOM tree prediction and fine-grained code synthesis. Specifically, we employ a ViT-based structural model to predict DOM trees, enhanced by a BBox-based data pruning strategy during training. Using the coarse DOM trees and bounding boxes generated by the structural model, we further simplify the code generation into two sub-steps: first generating code snippets for segmented subregions of the input image, then assembling these snippets into the final code. This decomposition effectively reduces the complexity associated with generating lengthy code sequences in a single step. Extensive experimental evaluations and feedback from human annotators confirm the effectiveness of our approach.


\balance
\bibliographystyle{ACM-Reference-Format}
\bibliography{ref}

\end{document}